\documentclass[12pt] {article}
\def\be{\begin{equation}}
\def\ee{\end{equation}}
\def\bea{\begin{eqnarray}}
\def\eea{\end{eqnarray}}
\def\pd{\partial}
\def\eps{\epsilon}
\def\vmu{\vec\mu}
\def\vnu{\vec\nu}
\def\vrho{\vec\rho}
\def\vkap{\vec\kappa}
\def\vkp{\vec\kappa'}
\def\vkpp{\vec\kappa''}

\def\vsig{\vec\sigma}
\def\vtau{\vec\tau}

\begin{document}
\title{Hamilton-Jacobi equations and Brane associated Lagrangians}
\author{L.M. Baker$\footnote{e-mail: l.m.baker@durham.ac.uk}$\,  and 
D.B. Fairlie$\footnote{e-mail: david.fairlie@durham.ac.uk}$\\
\\
Department of Mathematical Sciences,\\
         Science Laboratories,\\
         University of Durham,\\
         Durham, DH1 3LE, England}
\maketitle
\begin{abstract} This article seeks to relate a recent proposal for the association of a  covariant Field Theory with a string or brane Lagrangian to the Hamilton-Jacobi formalism for strings and branes. It turns out that since in this special case, the Hamiltonian depends only upon the momenta of the  Jacobi fields and not the fields themselves, it is the same as  a Lagrangian, subject to a constancy constraint. We find that the associated Lagrangians for strings or branes have a covariant
description in terms of the square root of the same Lagrangian. If the Hamilton-Jacobi function is zero, rather than a constant, then it is in in one dimension lower, reminiscent of the `holographic' idea. In the second part of the paper, we discuss properties of these Lagrangians, which 
 lead to what we have called `Universal Field Equations',  characteristic of covariant equations of motion.
\end{abstract}
 
\section{Introduction}
Towards the end of the last century, we proposed the association of a field theory with string theory and generalised this to branes \cite{bf}. This is not an entirely new idea, as 
something similar has been suggested by Hosotani \cite{hos1} and Morris \cite {morris1}\cite{morris2}, but their motivation was rather different. Our starting point was that of wave-particle duality in Quantum Theory, namely that a particle, described by the Lagrangian
\be
{\cal L}_1\,=\,\sqrt{\sum\left(\frac{\pd X^\mu}{\pd\tau}\right)^2}\label{one}
\ee
may also be described in  terms of a Klein Gordon field;
\be
{ L}_2\,=\,\ \frac{1}{2}{\sum\left(\frac{\pd \phi}{\pd x_\mu}\right)^2}\label{KG3}
\ee
(for a massless particle). 
Is there a similar alternative description of strings and branes?
We proposed that  a $D$-brane with Lagrangian \cite{born}\cite{dirac}
\be
\sqrt{\det\left|\frac{\pd X^\mu}{\pd\tau_i}\frac{\pd X^\mu}{\pd\tau_j}\right|}\label{bornbrane}
\ee
where where $\mu\,=\, 1\dots d$ and $i,j$ run over $D+1$ values
should also be  related to a Lagrangian for $D+1$ fields $\phi$
with Lagrangian
\be
{\cal L}=\sqrt{\det\left|\frac{\pd \phi^i}{\pd x_\mu}\frac{\pd \phi^j}{\pd x_\mu}\right|}.\label{inersebrane}
\ee
We shall refer to this as the {\it companion Lagrangian}, and the associated equations as {\it companion equations}.
Our proposal, like \cite{hos2}, involves $D+1$ fields in $d$ dimensions.  One unresolved issue in our earlier article was that we had no cogent argument for taking the square root in the  companion Lagrangian rather than the  Lagrangian without the square root, as the analogy with the Klein Gordon equation might suggest, apart from the additional bonus of the general covariance of the equations arising from the square root form and the idea that these Lagrangians form a natural continuation of the Dirac-Born-Infeld concept into the region where $d>D+1$.  Indeed, as we may illustrate for the cases $d=3,\ D=1$ and $d=2,\, D=2$ both Lagrangians may be expressed in the same form;
$${\cal L}\,=\,\sqrt{\sum_{p}\epsilon_{abijp}\epsilon_{cdklp}
\frac{\pd X_a}{\pd x_i}\frac{\pd X_b}{\pd x_j}\frac{\pd X_c}{\pd x_k}\frac{\pd X_d}{\pd x_l}},
$$
where the number of  derivatives of fields/co-ordinates is twice the lesser of $D+1$ and $d$ and the number of indices in the $\epsilon$ tensors is $D+1+d$,
$|d-D-1|$ of which are contracted, in the general case.
 
The original proposal of Hosotani and, independently, Morris was to associate with the
string, or more generally $D$-brane, a Lagrangian of the same form as (\ref{inersebrane}) but now with the complementary number of fields, $d-D-1$.
In their case the Lagrangian, up to a Jacobian factor is obtained from (\ref{inersebrane}) but now with the complementary number of fields, $d-D-1$.
In their case the Lagrangian, up to a Jacobian factor is obtained from (\ref{bornbrane}) by exchanging the r\^oles of dependent and independent variables, and the resulting equations of motion are classically equivalent
to those of the original Lagrangian. This is the reason for taking the complementary number of fields.  As we shall see, the number of independent equations arising from (\ref{bornbrane}) is, as a result of re-parametrisation
invariance not $d$, but $d-D-1$, which precisely matches up with those from
the Hosotani/Morris procedure and  furthermore the equations themselves transform into each other under the exchange of dependent and independent variables, so that there is a complete mathematical equivalence at the classical level between the two formalisms. In a later paper, Hosotani and Nakayama \cite{hos2} have advocated instead a Hamilton-Jacobi formulation for classical strings and branes, developing some ideas in \cite{hos1}  which rely upon formulations of Hamilton-Jacobi theory for many fields developed by Rund,  Nambu and Kastrup \cite{Rund} \cite{nambu}\cite{kastrup}, based upon pioneering work by Carath\'edory \cite{car} and Velte  \cite{velte}. Nambu's approach differs from the others in an important regard. Essentially, his Hamilton-Jacobi function for strings is the square of that for Kastrup and Rinke. We want to take this function as equivalent to a Lagrangian, since in this particular case there is no explicit dependence upon the fields,but only their derivatives.    We shall show, using the quadratic form of the H-J function as Lagrangian, subject to the constraint that it is constant gives the same equations of motion  as that for a Lagrangian which is the square root of the H-J function, now unconstrained. Furthermore if this constant is zero then together with additional constraints  permit the elimination of the dependence upon a designated co-ordinate the theory is equivalent to that from a Lagrangian equivalent to the square root of this H-J function in one dimension less, but where it is now unconstrained. In either case the new Lagrangian has a similar form to a continuation of the Dirac-Born-Infeld Lagrangian to the case where the dimensions of the base space exceed those of the target space.

Note that this association of a field theory with a string is contrary to the standard received wisdom that such a field should be a functional of the arc length parameter.  However the idea seems such a natural extension of the notion of the Dirac-Born-Infeld Lagrangians that it seems to us worth pursuing. 
\section{ Hamilton-Jacobi  and the Companion Lagrangian}
This section begins with a paraphrase of Nambu's work \cite{nambu},  He bases this upon the idea of extending the one-form relation
\be
dS\,=\,\sum_i p_idx_i -H dt,\ \ H\,=\,H(p_i,x_i),\ \ S\,=\, S(x_i,t).\label{onef}
\ee
He replaces this with a two-form relation
\be
dS_0\wedge dT_0 +dS\wedge dT\,=\,\sum_{i>j}p_{ij}dx_i\wedge dx_j -Hd\sigma\wedge d\tau.\label{twof}
\ee
Here $H$ is a function  $H(p_{ij},x_k)$, and, in virtue of the fact that there are no cross terms between $dx_j$  and $d\sigma,\ d\tau$, $S_0,\ T_0$ may be taken as functions only of $\sigma,\ \tau$, while $S,\ T $ depend only upon the co-ordinates $x_j$. Then 
\bea
p_{ij}&=&\frac{\pd(S,\ T)}{\pd(x_i,\ x_j)}\label{jac1}\\
H&=&\frac{\pd(S_0,\ T_0)}{\pd(\sigma,\ \tau)}\label{jac2}
\eea
The analogues of the usual Hamiltonian equations become
\be
\frac{\pd(x_i,\ x_j)}{\pd(\sigma,\ \tau)}\,=\,\frac{\pd H}{\pd p_{ij}},\ \ \sum_j\frac{\pd(p_{ij},\ x_j)}{\pd(\sigma,\ \tau)}\,=\,-\frac{\pd H}{\pd x_{i}}
\label{analog}
\ee
These equations imply 
$$
\frac{\pd H}{\pd \sigma}\,=\,0\ \ {\rm and}\ \ \frac{\pd H}{\pd \tau}\,=\,0.
$$
This means that $H$ is a constant of the motion, and does not depend explicitly upon the evolution parameters.
In the case of the Schild action \cite{schild}\cite{eguchi}, the Hamiltonian is
equivalent to the Lagrangian as there is no dependence upon the fields, but only upon their derivatives and may be written as
\be
H\,=\,\frac{1}{2}\sum_{\mu>\nu} p_{\mu\nu}^2\,=\,-{\cal L}\,=\, {\rm constant}
\label{schild}
\ee
and the equations of motion are the same as  for the Nambu-Goto action;
\be
\sum_j\frac{\pd(p_{ij},\ x_j)}{\pd(\sigma,\ \tau)}\,=\,0. \label{eqmo}
\ee
The other equations of motion are
\be
p_{ij} \,=\,\frac{\pd(x_i,x_j)}{\pd(\sigma,\ \tau)}\label{jac3}
\ee
Hosotani and  Nakayama \cite{hos2} observed that equations (\ref{eqmo}) when the momentum tensors $p_{ij}$ are regarded as functions of the $x_k$ may be written as
\be
p^{ij}\frac{\pd p_{jk}}{\pd x_i}\,=\,0,\label{secret}
\ee
and these equation, with the help of (\ref{jac1}) may be used to show that
$p^{ij}p_{ij}$ is constant, which is consistent with (\ref{schild}).

We now make the following  observation; the usual Hamilton-Jacobi relation for a massless point particle,
$$ \left(\frac{\pd S}{\pd x_\mu}\right)^2 \,=\,0$$
takes the same form as the Klein Gordon Lagrangian for a massless particle subject to a constraint. What happens when we subject the equation of motion for the  Lagrangian to  the constraint that the Lagrangian is zero?
Consider the Lagrangian for the Klein-Gordon equation in $d$ space-time dimensions.
\be
{\cal L}=\left(\frac{\pd \phi}{\pd x^{\mu}}\right)^2 \qquad \textrm{where} \qquad \mu=1,\ldots, d \label{L1}
\ee
By imposing the condition ${\cal L}=M^2$ where $M$ is constant the equations of motion for the above Lagrangian are now the same as the equations of motion obtained from the square root of the unconstrained Lagrangian since they may be written as
\be 
\frac{\pd^2 \phi}{\pd x_\mu^2}-\frac{\pd \phi}{\pd x_\mu}\frac{\pd\sqrt{{\cal L}}}{\pd x_\mu}\,=\,\frac{\pd \phi}{\pd x_\mu}\frac{\pd\sqrt{{\cal L}}}{\pd x_\mu}\label{addsub}
\ee
where the same (zero) quantity has been subtracted from both sides. The vanishing of the left hand side is just the equation of motion for the square root of  ${\cal L}$. If the constant $M$ is zero then we can go further and assert that the equations of motion are the same as those obtained fom the square root of the unconstrained Lagrangian in one dimension less,i.e.

\be
{\cal L}=\sqrt{\left(\frac{\pd \phi}{\pd x^{\mu}}\right)^2}\qquad  \textrm{where}\qquad  \mu=1,\ldots, d-1.\label{L2}
\ee
Remarkably, this procedure generalizes. In the case of the string, the Hamilton Jacobi equation is equivalent to the Lagrangian
\be
\frac{1}{2}\sum_{\mu>\nu} p_{\mu\nu}^2\,=\,-{\cal L}\,=\left(\frac{\pd S}{\pd x_\mu}\right)^2\left(\frac{\pd T}{\pd x_\nu}\right)^2-\left(\frac{\pd S}{\pd x_\mu}\frac{\pd T}{\pd x_\mu}\right)^2.\label{lag}
\ee
Here $\mu,\ \nu\,=\, 1\dots d$. It is necessary to impose something more stringent than ${\cal L}=$ constant, which is the Hamilton Jacobi equation. We want to have
\be \delta{\cal L} \,=\, \frac{\pd {\cal L}}{\pd S_\nu}\delta S_\nu\,+\,\frac{\pd {\cal L}}{\pd T_\nu}\delta T_\nu\,=\,0.\label{var}
\ee
The usual assumption is to treat the variation in the derivative $\frac{\pd S}{\pd x_\nu}$ and that in $\frac{\pd T}{\pd x_\nu}$ as connected, but, suppose we treat these variations as independent;

 \bea 
 \frac{\pd {\cal L}}{\pd S_\nu}S_{\nu\mu}&=&0,
\label{cons1}\\
 \frac{\pd {\cal L}}{\pd T_\nu} T_{\nu\mu}&=&0,
\label{cons2}
\eea

  In fact these constraints are not fully  independent, as the sum of each pair
 is equivalent to the equations (\ref{secret}) or $\displaystyle{\frac{\pd{\cal L}}{\pd x_\mu}\,=\,0}$ .
Together, they permit the elimination of all second order derivatives $\displaystyle{\frac{\pd^2 S}{\pd x_d\pd x_\mu},\  \frac{\pd^2 T}{\pd x_d\pd x_\mu}}$ where one or more of the derivatives is with respect to $x_d$ from the equations of motion for ({\ref{lag}}). The constraints are obtained by differentiating ${\cal L}$ with respect to $x_\mu$ first regarding $T_\mu$ as constant, then repeating the procedure, now regarding $S_\mu$ as constant, etc and take the generic form
$$ \frac{\pd{\cal L}}{\pd\frac{\pd\phi^i}{\pd x_\mu}}\frac{\pd^2\phi^i}{\pd x_\mu\pd x_\nu}\,=\,0,\ \ i\ {\rm not\ \ summed}.
$$
The resulting equations are precisely the equations of motion coming from the Lagrangian $\sqrt{{\cal L}}$, but in one dimension less, i.e. $d-1$ dimensions.
It is also necessary to take the constant to be zero. Notice also that if the
$d=D+2$ then the equations of motion for the reduced Lagrangian vanish identically, as the square root Lagrangian is now proportional to the Jacobian of the fields, and is thus a divergence. 
In contrast to the particle case, where these assertions are
readily verified, the proof is rather tricky. Computer calculations have verified this conjecture for the cases $D\,=\,2$, $d\,=\, 3,\ 4,\ 5$ and $D\,=\, 3,\ d\,=\,4,\ 5$. This encouraged us to seek an analytic proof, given first for the case of two fields, and  established in general by one of us, Linda Baker, while this paper has been under revision. Both these proofs are contained in appendices.  Thus the procedure with which we proposed to associate a field theory with a string or Brane Lagrangian gains some justification from the Hamilton-Jacobi approach, with the added refinement that if the Hamilton-Jacobi function is zero, the  Lagrangian is defined in a space of one dimension less than that of the target space of the string. This property
calls to mind the Holographic Principle of 't Hooft \cite{hooft}, developed by Susskind \cite{susskind}, where the theory in the bulk is determined by the 
theory on the boundary, i.e in one dimension less. 
There is a second issue which suggests that there exists a more general class of constraints than those proposed above which  also result in the equivalence of the equations of motion to those from a square root Lagrangian in one lower dimension. The Lagrangian (\ref{lag}) is invariant under canonical transformations of $S,T$, and in particular is invariant under rotations of these fields, whereas the constraints (\ref{cons1},\ref{cons2}) are not. While these constraints fulfil the purpose for which they were contrived, this observation suggests that they are sufficient, but not necessary. 

\section{Properties of square-root Lagrangians.}

It turns out that actions constructed from Lagrangians of the form (\ref{inersebrane})
 possess remarkable properties, just as do those of Dirac-Born-Infeld type (\ref{bornbrane}). The latter are invariant under re-parametrisation (diffeomorphism and dilation) invariance, whereas the field theory Lagrangians (\ref{inersebrane}) are generally covariant. This property is reminiscent of the
construction of what we have called `Universal Field Equations' \cite{gov}. One purpose of this paper is to show that indeed there is a connection, and that the equations which
arise are universal in the sense that they are the Euler-Lagrange equations for a whole class of inequivalent Lagrangians. Furthermore, there is evidence for 
universality associated with a  second variation, using an iterated form of Lagrangian.  To explain in detail what we mean, we consider first of all the case of a single field, associated with the point particle. Consider the square root of the Klein Gordon Lagrangian;
\be
{\cal L}\,=\,\ \sqrt{{\sum\left(\frac{\pd \phi}{\pd x_\mu}\right)^2}},\label{KG4}
\ee
 As remarked in \cite{bf}, in the  minimal case of two base co-ordinates the equation of motion is the well known Bateman equation \cite{gov}
\be
\left(\frac{\pd \phi}{\pd x_1}\right)^2\frac{\pd^2 \phi}{\pd x_2^2}+
\left(\frac{\pd \phi}{\pd x_2}\right)^2\frac{\pd^2 \phi}{\pd x_1^2}-
2\left(\frac{\pd \phi}{\pd x_1}\right)\left(\frac{\pd \phi}{\pd x_2}\right)\frac{\pd^2 \phi}
{\pd x_1\pd x_2}\,=\,0,\label{batman}
\ee
This is the simplest example of a Universal Field Equation, since it arises as the variation of any Lagrangian $\displaystyle{{\cal L}\left(\phi,\ \frac{\pd \phi}{\pd x_1},\ \frac{\pd \phi}{\pd x_2}\right)}$, which is homogeneous of weight one in the first derivatives; i.e.
\be
\frac{\pd \phi}{\pd x_1}\left(\frac{\pd{\cal L}}{\pd \frac{\pd \phi}{\pd x_1}}\right)\,+\,\frac{\pd \phi}{\pd  x_2}\left(\frac{\pd{\cal L}}{\pd \frac{\pd \phi}{\pd x_2}}\right)\,=\,{\cal L}\label{hmog}
\ee
All such Lagrangians which are independent of $\phi$ are clearly independent under field redefinitions, since any such transformation from a $\phi$-independent Lagrangian will inevitably introduce a field. If we now consider the base space in general dimension, with such a Lagrangian of weight one, the equations of motion are sums of Bateman equations for all pairs of independent variables. This follows simply from the fact that the equation of motion is simply (denoting derivatives by subscripts);
\be
\frac{\pd^2{\cal L}}{\pd\phi_\mu\pd\phi_\nu}\phi_{\mu\nu}\,=\,0,\label{eqm}
\ee
and the homogeneity condition implies that 
\be
\sum_\mu \phi_\mu\frac{\pd^2{\cal L}}{\pd\phi_\mu\pd\phi_\nu}\,=\,0.\label{hom2}
\ee
These equations can be solved for the double derivatives $\frac{\pd^2{\cal L}}{\pd\phi_\mu^2}$ in favour of the mixed derivatives, and the equation of motion with $d$ independent variables becomes a sum of Bateman equations;
\be
\sum_\mu\sum_{\nu<\nu}\frac{1}{\phi_\mu\phi_\nu} \frac{\pd^2{\cal L}}{\pd\phi_\mu\pd\phi_\nu}\left(\left( \phi_\mu\right)^2\phi_{\nu\nu}+
\left(\phi_\nu\right)^2 \phi_{\mu\mu}-
2\left(\phi_\mu\phi_\nu\right)\phi_{\mu\nu}\right)\,=\,0,\label{eqn2}
\ee
Now, as is well known, the solution of any individual Bateman equation (\ref{batman}) is given implicitly by solving for $\phi$ the equation 
\be
x_\mu F(\phi(x_1,x_2\dots x_d))\,+\,x_\nu G(\phi(x_1,x_2\dots x_d))\,=\,c,
\ee
where $F,\ G$ are arbitrary functions and $c$ is a constant. A simple analysis
shows that a large class of universal solutions (in the sense that they are independent of details of the Lagrangian)  of (\ref{eqn2}) will be given by solving the implicit equation
\be
\sum_{\mu=1}^{\mu=d} x_\mu F_\mu(\phi(x_1,x_2\dots x_d))\,=\,c,\label {univsol}
\ee
where $F_\mu$ are arbitrary functions of $\phi$. In particular, it provides a solution to the equations of motion arising from (\ref{KG4}).

This class of solutions is also a solution of the Universal Field Equation, obtained by iterating the Euler operator ${\cal E}$ acting on the Lagrangian
\be
{\cal E}=-\frac{\pd}{\pd\phi}
 +\pd_i \frac{\pd}{\pd\phi_i}-\pd_i\pd_j\frac{\pd}{\pd\phi_{ij}}\dots
\label{elop}
\ee
(In principle the expansion continues indefinitely  but it is sufficient for
our purposes to terminate at the stage of second derivatives  $\phi_{ij}$,
since it turns out that the iterations do not introduce any derivatives
higher than the second).

Then the $d-1$ fold iteration
\be
 {\cal E}{ \cal L} {\cal E}{\cal L}{\cal E}{\cal L },\cdots,{\cal E}{\cal L}\label{iter}
\ee
where each Euler operator acts on everything to the right yields the Universal
Field Equation
\be
\det\pmatrix{0&\phi_1&\phi_2&\ldots&\phi_d\cr
               \phi_1&\phi_{11}&\phi_{12}&\ldots&\phi_{1d}\cr
               \phi_2&\phi_{12}&\phi_{22}&\ldots&\phi_{2d}\cr
                    .&\       .&\       .&\ddots&\       .\cr
               \phi_d&\phi_{1d}&\phi_{2d}&\ldots&\phi_{dd}\cr}\,=\,0,
\label{univ}
\ee
This is proved in \cite{gov}. This equation, as has been remarked above, admits a class of solutions of the form (\ref{univsol}).
Now let us turn to the 
properties of the particle Lagrangian, (\ref{bornbrane}). Here there are  $d-1$ independent equations of motion
 which take the form
\be
\frac{\pd^2 X^{\mu}}{\pd \tau^2} \frac{\pd X^{\nu}}{\pd \tau}-\frac{\pd^2 X^{\nu}}{\pd \tau^2} \frac{\pd X^{\mu}}{\pd \tau}=0.\label{class1}
\ee
The reason that there are not $d$ independent equations is due to\\ re-parametrisation invariance. Furthermore, exchanging the roles of dependent and independent variables each such  equation  reproduces the Bateman equation, with independent variables $x_\mu,\ x_\nu$. 
Thus there is a direct connection between solutions of (\ref{bornbrane}) and a large class of solutions to the companion equation(\ref{inersebrane}). In what follows we elaborate upon this situation for the case of more fields, and find that the appropriate variables are Jacobians of the fields.
\section{Companion equations for the string}
 For  the Nambu-Goto Lagrangian,
\be
{\cal L}_{d=2}\,=\,\sqrt{\sum\left[\left(\frac{\pd X^\mu}{\pd\sigma}\frac{\pd X^\mu}
{\pd\tau}\right)^2-\left(\frac{\pd X^\mu}{\pd\sigma}\right)^2\left(\frac{\pd X^\mu}{\pd\tau}\right)^2\right]}\label{three}
\ee
describing strings in $d$ dimensions there are $d-2$ independent equations of motion. 
 For example, for the  string case in d=3 dimensions, the Nambu-Goto Lagrangian ${\cal L}_3$ gives the single equation of motion
\be
\left( \begin{array}{ccc}
\hat{J}_1 & \hat{J}_2 & \hat{J}_3
\end{array} \right) 
\left( \begin{array}{ccc}
X^1_{\sigma\sigma} & X^1_{\sigma\tau} & X^1_{\tau\tau}\\
X^2_{\sigma\sigma} & X^2_{\sigma\tau} & X^2_{\tau\tau}\\
X^3_{\sigma\sigma} & X^3_{\sigma\tau} & X^3_{\tau\tau} 
\end{array} \right)  
\left(  \begin{array}{c}
(X^1_\tau)^2+(X^2_\tau)^2+(X^3_\tau)^2\\
-2(X^1_\sigma X^1_\tau+X^2_\sigma X^2_\tau+X^3_\sigma X^3_\tau)\\
(X^1_\sigma)^2+(X^2_\sigma)^2+(X^3_\sigma)^2
\end{array} \right) =0\label{class2}
\ee
where
\be
X^\mu_{ij}=\frac{\pd^2 X^{\mu}}{\pd \sigma^i \pd \sigma^j},\qquad  X^\mu_i=\frac{\pd X^\mu}{\pd \sigma^i}, \qquad \textrm{and}\quad \sigma^i=(\sigma, \tau), \label{class3}
\ee
and
\be
\hat{J}_\rho=\epsilon_{\rho\mu\nu} X^\mu_\sigma X^\nu_\tau =\frac{1}{2} \epsilon_{\rho\mu\nu}
\left| \begin{array}{cc}
X^\mu_\sigma  & X^\nu_\sigma\\
X^\mu_\tau & X^\nu_\tau
\end{array} \right|.\label{class4}
\ee
In general, a typical equation of motion, of which only $d-2$ are independent can be written in the following form:
\be
\hat{J}_\nu X^\nu_{ij} (L^{-1})_{ij} = 0,\label{class5}
\ee
where L is the matrix with components $[L]_{ij}=\frac{\pd X^\mu}{\pd\sigma^i}\frac{\pd X^\mu}{\pd \sigma^j}$ and $\nu$  is chosen from three of the values $\nu_1,\ \nu_2,\ \nu_3$  of the index $\mu$ which runs over $1\dots d$. $\hat J_{\nu_1}$ denotes the Jacobian $\displaystyle{\frac{\pd( X^{\nu_2}, X^{\nu_3})}{\pd(\ \sigma_1,\ \sigma_2\ )}}$, omitting $X^{\nu_1}$ etc.
 This can be extended to strings in $d$ dimensions and to branes. The only essential difference is that in the typical equation of motion, $\nu$ is now an arbitrary choice of $D$ values and $\hat J_\nu$ is now a Jacobian of a subset  of those  variables
$x^\nu$, with respect to the $d$ world sheet co-ordinates
$\sigma_j$.

In general, an object (particle/string/brane)  which sweeps out an $N$-dimensional world volume in $d$-dimensional space-time has only $d-N$ independent equations of motion. The basic reason for this is that in the case $N\,=\,d$ the Lagrangian is a divergence, so all the equations
of motion vanish. The companion Lagrangian, suggested in \cite{bf}
is
\be
{\cal  L}_{N=2}\sqrt{\sum\left[\left( \phi_\mu\psi_\mu\right)^2-
\left(\phi_\mu\right)^2\left( \psi_\nu\right)^2\right]}
\,=\, -\frac{1}{2}\sqrt{\sum_{\mu,\ \nu}\left ( \phi_\mu \psi_\nu-\phi_\nu \psi_\mu\right)^2}
\label{four}
\ee
Note that this Lagrangian depends upon the derivatives of $\phi,\ \psi$ only through the Jacobians $J^{\mu\nu}\,=\,\phi_\mu \psi_\nu-\phi_\nu \psi_\mu$, as it may be written in the second alternative form. This is a clue as to the universal properties of such Lagrangians. In the case where $\mu\,=\,1\ldots3$,
using the notation $j^\mu\,=\, \frac{1}{2}\epsilon_{\mu\nu\rho}J^{\nu\rho}$, an easy calculation shows that any Lagrangian homogeneous of degree one in the variables $j^\mu$ will give the same equations of motion \cite{gov2}\cite{dbf}
\cite{rev}
\be
\det\left|\begin{array}{ccccc}0&0&\phi_{x_1}&\phi_{x_2}&\phi_{x_3}\\
                         0&0&\psi_{x_1}&\psi_{x_2}&\psi_{x_3} \\                                                                      \phi_{x_1}&\psi_{x_1}&\phi_{x_1x_1}&\phi_{x_1x_2}&\phi_{x_1x_3}\\
\phi_{x_2}&\psi_{x_2}&\phi_{x_1x_2}&\phi_{x_2x_2}&\phi_{x_2x_3}\\
\phi_{x_3}&\psi_{x_3}&\phi_{x_1x_3}&\phi_{x_2x_3}&\phi_{x_3x_3}\end{array}\right|\,=\,0.
\label{3d}
\ee
Furthermore, just as in the case of one field, the companion Lagrangian to the 
Brane, (\ref{inersebrane}) in any number of dimensions,  is expressible as the square root of a sum of squares, now of Jacobians,
\be
{\cal L}\,=\,\sqrt{\det\left|\frac{\pd \phi^i}{\pd x_\mu}\frac{\pd \phi^j}{\pd x_\mu}\right|}\,=\, \left(\frac{N!(d-N)!}{d!}\right)\sqrt{\sum\left(\frac{\pd\{\phi^1,\ \phi^2,\dots,\phi^n\}}{\pd\{x_{\mu_1},x_{\mu_2}\dots x_{\mu_d}\}}\right)^2}.\label{field}
\ee
(Here the sum is over all permutations of the squares of Jacobians of the $N$ fields  with respect to  selections of $N$ (the dimension of the world volume) out of the $d$ co-ordinates $x_\mu$ of space-time).
If it is substituted by any homogeneous function of these Jacobians of weight one, the Lagrangian will give give equations of motion which
take the form of a weighted sum of Universal Field Equations of the form of (\ref{3d}), suitably generalised to $N$ fields. As is the case with a single field, a class of solutions may be founfd by taking the equations
\bea
\sum_{\mu=1}^{\mu=d} x_\mu F_\mu(\phi(x_1,x_2\dots x_d))&=&\,c_1,\label {univsol1}\\
\sum_{\mu=1}^{\mu=d} x_\mu G_\mu(\psi(x_1,x_2\dots x_d))&=&\,c_2,\label {univsol2}\\
\eea
where $F_\mu(\phi),\  G_\mu(\psi)$ are arbitrary functions, subject to the above constraints. This  class of solutions is more genenral than might be at first supposed, thanks to general covariance which asserts that any function of these
implicit solutions for $\phi$ and $\psi$
 is also a solution.
 There will be only one contribution in the case $d\,=\,D+1$, the multi-field Universal Equation of \cite{gov} and \cite{rev}. The equation of motion in this case is completely classically equivalent to the single string equation, as may be seen by inverting the r\^oles of dependent and independent variables. This  observation suggests that
the received wisdom, that string Field theory should depend upon a field which is a functional of the string, may be unnecessary and a quantised version of 
the companion Lagrangian sufficient. Of course this is a highly non-linear theory
whose quantisation is problematic.
\subsection{Inclusion of a Background Metric}
The property of the companion Lagrangian, and also of Lagrangians of the Born-Infeld type that they may be expressed in terms a quadratic form in Jacobians persists even when a background metric $g_{\mu,\nu}$ is included. For example in the case of
the string companion field, with
\be
{\cal L} = \sqrt{\det\left|g_{\mu\nu}\frac{\pd \phi_i}{\pd x_\mu}\frac{\pd \phi_j}{\pd x_\nu}\right|}\label{gcomp}
\ee
the Lagrangian may be expressed as 
$$ 
{\cal L} = \sqrt{\sum_{\mu>\nu} ( g_{\mu\nu}g_{\rho\sigma}- g_{\mu\sigma}g_{\rho\nu})\left(\frac{\pd (\phi_1,\phi_2)}{\pd (x_\mu x_\nu)}\right)\left(\frac{\pd (\phi_1,\phi_2)}{\pd (x_\rho x_\sigma)}\right)} .
$$
The pattern for more fields is similar involving the minors of $g$ of rank
$D+1$.

\section{Iterated Variations}
In this section, which is by way of a speculative generalisation of the iterative construction \cite{gov} outlined at the end of section 5, we shall discuss the extension of the scheme  for the construction of a sequence of iterated Lagrangians to ones which depend on more than one field. In the introduction the iterative procedure for Lagrangians for one field $\phi$ just involved multiplying by the Lagrangian before re-applying the Euler operator. It is not quite so simple for Lagrangians of two fields such as (\ref{four}). 
Consider Lagrangians which are homogeneous of weight one in the Jacobians. From now on ${\cal E}_\phi$ denotes the Euler operator with respect to field $\phi$. Carrying out the iteration (\ref{iter}) now results in derivatives of orders higher than two (unlike the one field case where all higher order derivatives vanished). However, this problem is removed if we multiply by a function $f$ which is related to the Lagrangian and which is homogeneous of weight one in $\phi_\mu$ and weight zero in $\psi_\mu$ so that overall $f {\cal E}_\phi{\cal L}$ is weight one in both $\phi_\mu$ and $\psi_\mu$, like the Lagrangian. 
In order to keep derivatives of order no higher than two, the function $f$ must satisfy the conditions
\bea
\frac{\frac{\pd f}{\pd \phi_1}}{\frac{\pd f}{\pd \psi_1}}=\frac{\frac{\pd f}{\pd \phi_2}}{\frac{\pd f}{\pd \psi_2}}=\frac{\frac{\pd f}{\pd \phi_3}}{\frac{\pd f}{\pd \psi_3}}=\frac{\frac{\pd^2 {\cal L}}{\pd \phi_1 \pd\phi_1}}{\frac{\pd^2 {\cal L}}{\pd \psi_1 \pd \phi_1}}
\eea
 Consider the case where $d=3$ and the Lagrangian is (\ref{four}). The required function $f$ was found to be 
\be
f=\frac{{\cal L}}{\sqrt{\psi_1^2+\psi_2^2+\psi_3^2}}
\ee
The iterative sequence is then
\be
{\cal E}_\phi f {\cal E}_\phi{\cal L}=\frac{1}{(\psi_1^2+\psi_2^2+\psi_3^2)^{3/2}} 
\det\left|\begin{array}{cccc}
 0 & \psi_1&\psi_2&\psi_3\\
\psi_1 & \psi_{11}&\psi_{12}&\psi_{13}\\
\psi_2 & \psi_{12}&\psi_{22}&\psi_{23}\\
\psi_3 & \psi_{13}&\psi_{23}&\psi_{33}\end{array}\right|\label{efel}
\ee
The following list of Lagrangians all behave similarly and suitable $f$'s have been found for them all.
\bea
&&{\cal L}=\sqrt{a_{ij} J_i J_j}, \qquad \qquad f=\frac{{\cal L}}{\sqrt{2 \epsilon_{i_1i_2i_3}\epsilon_{j_1j_2j_3} a_{i_2j_2} a_{i_3j_3} \psi_{i_1}\psi_{j_1}}}\\
&&{\cal L}=\frac{a_{ij} J_i J_j}{c_k J_k},\qquad \qquad \;\; f=\frac{c_k J_k}{\sqrt{\epsilon_{i_1i_2i_3}\epsilon_{j_1j_2j_3} a_{i_2j_2} c_{i_3}c_{j_3} \psi_{i_1}\psi_{j_1}}}\\
&&{\cal L}={\cal L}(b_k J_k, c_k J_k), \qquad f=\frac{{\cal L}}{\epsilon_{ijk} c_i b_j \psi_k}
\eea 
The $a_{ij}, b_k, c_k$ are all constants and summation over indices is assumed. The first of these examples is just the case of (\ref{four}) in a background metric.  
In all the above cases the iterated sequence always results in the same form as in (\ref{efel}). The determinant part always appears and is multiplied by some factor. The determinant is a generalised Bateman equation. It is not particularly surprising that the second iteration of these Lagrangians is the same, but what is important is that it only depends on the first and second derivatives of $\psi$ and has no dependence on $\phi$ at all. This is analogous to all $\phi$ dependence disappearing in the one field case after two iterations.

Similar functions  $f$ can be found for  other cases where the number of fields is one less than the number of dimensions . It is hoped to extend the results to Lagrangians with $D$ fields in $d$ dimensions.
\section{Conclusions}
The main result of this paper is that there is a natural continuation of the Dirac-Born-Infeld Lagrangian to the case where the number of base space co-ordinates exceeds the number of target space ones, and that this Lagrangian follows from the Hamilton-Jacobi equation for Strings and Branes. Two arguments have been advanced that these Lagrangians should be considered as the square of  a quadratic in Jacobians, rather than the quadratic form itself; firstly on the grounds of covariance and secondly as an interpretation of the Hamilton-Jacobi function for Strings and Branes as a Lagranian constrained to be constant.  When the Lagrangian is subject to a number of vanishing constraints, it gives rise to equations of motion identical with those from an unconstrained Lagrangian in one dimension less. 
This may be interpreted as a simple example of the Holographic principle, as the
evolution of the system is determined by equations on the boundary. Obviously
this notion gives rise to many questions for further development, such as 
the extension to include an Abelian field, and the further understanding of the additional constraints besides the vanishing of the Lagrangian on the space of 
solutions of the equations of motion, a property we have referred to as pseudo-topological \cite{bf}\cite{dbf}. There are also questions of universality
which arise, such as the determination of the class of Lagrangians for which our result holds.

\section*{Acknowledgement}
Linda Baker is grateful to EPSRC for a postgraduate research award.
\newpage
\section*{APPENDIX}
For ease of understanding, we present the proof, first for two fields, then for an arbitrary number of fields.
\subsection*{Part 1}
   This contains a proof of the theorem that equations of motion arising from the quadratic Lagrangian
\be
{\cal L}\, =\, S_\mu^2 T_\nu^2 -(S_\mu T_\mu)^2\, =\, (S\cdot S)( T\cdot T) - (S\cdot T)^2 \label{again}
\ee
subject to the constraints
\be
{\cal L}\,=\,0,\ \  \frac{\pd {\cal L}}{\pd S_\nu}S_{\mu\nu}\,=\,0,\ \
\frac{\pd {\cal L}}{\pd T_\nu}T_{\mu\nu}\,=\,0,\label{restrict}
\ee 
are exactly equivalent to the equations arising from the Lagrangian of similar
form, but with the square root taken, viz; 
\be
{\cal L}\, =\, \sqrt{S_i^2 T_j^2 -(S_j T_j)^2},\label{rootl}
\ee
in one dimension less. (Greek indices run from 1 to $d$, Roman from 1 to
$d-1$).
The equation for the field $S$ arising from the two equations of motion for
(\ref{again}) is , after elimination of second derivatives $S_{d\mu}$ takes the 
form
$$\left(\left(\frac{\pd {\cal L}}{\pd S_d}\right)^2\frac{\pd {\cal L}^2}{\pd S_iS_j}
-\frac{\pd {\cal L}}{\pd S_d}\frac{\pd {\cal L}}{\pd S_i}\frac{\pd {\cal L}^2}{\pd S_dS_j}-\frac{\pd {\cal L}}{\pd S_d}\frac{\pd {\cal L}}{\pd S_j}\frac{\pd {\cal L}^2}{\pd S_dS_i}+\frac{\pd {\cal L}}{\pd S_i}\frac{\pd {\cal L}}{\pd S_j}\frac{\pd {\cal L}^2}{\pd S_d^2}\right)S_{ij}=0$$
The coefficient of $S_{ij}$ may be expressed in the form of a determinant;
$$
\det\left|\begin{array}{ccc}
0&S_d(T\cdot T)-T_d(S\cdot T)&S_i(T\cdot T)-T_i(S\cdot T)\\
S_d(T\cdot T)-T_d(S\cdot T)&T\cdot T-T_d^2&-T_iT_d\\
S_j(T\cdot T)-T_j(S\cdot T)&-T_jT_d &\delta_{ij}(T\cdot T)-T_iT_j
\end{array}\right|
$$
 Using elementary row and column operations, this is equal to
$$
\frac{T\cdot T}{T_iT_j}\det\left|\begin{array}{ccc}
0&J_{di}&S_i(T\cdot T)-T_i(S\cdot T)\\
(T\cdot T)J_{dj}&T_iT_j+\delta_{ij}T_d^2&-\delta_{ij}(T\cdot T)T_d\\
S_j(T\cdot T)-T_j(S\cdot T)&-\delta_{ij}T_d &\delta_{ij}(T\cdot T)-T_iT_j.
\end{array}\right|
$$
Here $J_{di}$ denotes the Jacobian $J_{di}\,=\, S_dT_i-S_iT_d$.
Next we note that the product $(S_i(T\cdot T)-T_i(S\cdot T))(S_j(T\cdot T)-T_j(S\cdot T)$ can be written using ${\cal L}=0$ in the form
$\sum_\mu J_{i\mu}J_{j\mu}(T\cdot T)$.
Expanding the determinants, it is easy to obtain the coefficient of $\delta_{ij}$ from the first form, and of $T_iT_j$ from the second
giving the expression
$$
\left(T\cdot T\right)^2\left(-\sum_k J_{ik}J_{jk}-\delta_{ij}\sum J_{d\mu}^2\right)
$$
for the equation of motion.
For $i\neq j$ this is readily seen as the correct coefficient of $S_{ij}$ in the equation of motion for (\ref{rootl}). For $i=j$ it is also seen to be the correct coefficient,
with the factor $\displaystyle{\sum_{\stackrel{k,l\neq j}{k>l}}J_{kl}^2}$ using 
$${\cal L}\, =\,\sum_kJ_{jk}^2 +\sum_\mu J_{d\mu}^2+\sum_{\stackrel{k,l\neq j}{k>l}}J_{kl}^2\,=\,0.
$$
 
\subsection*{Part2. General Proof$\footnote{by Linda Baker}$}
 Here we give an analytic proof of this observation for any number of fields $n$ in any number of dimensions $d$ where $d>n$.

\subsection*{Conventions and Notation}
Partial derivatives are denoted by
\be
\frac{\pd\phi^i}{\pd x_\mu}=\phi^i_\mu,\qquad \frac{\pd^2\phi^i}{\pd x_\mu \pd x_\nu}=\phi^i_{\mu\nu}
\ee
Totally antisymmetric tensors $\eps_{\nu_1\nu_2\ldots\nu_d}$ are used throughout the proof with the convention that $\eps_{12\ldots d}=+1$. When indices have an arrow above them then they represent several indices. They can be thought of as vectors with several components.\\
$\vmu,\vnu,\vrho$ each have $(n-1)$ components. For example $\vmu$ denotes $\{\mu_2, \mu_3,\ldots,\mu_n\}$. \\
$\vtau, \vkap$ each have $(d-n)$ components. For example $\vkap$ denotes $\{\kappa_1, \kappa_2,\ldots,\kappa_{d-n}\}$.\\
$\vkp$ denotes $\{\kappa_2,\kappa_3,\ldots,\kappa_{d-n}\}$ and $\vkpp$ denotes $\{\kappa_3,\ldots,\kappa_{d-n}\}$.\\
For the product of $(n-1)$ fields we use the notation $\Phi_{\vnu}=\phi^2_{\nu_2}\phi^3_{\nu_3}\ldots\phi^n_{\nu_n}$.\\[2mm]
A useful identity which will be used later on is
\bea
\eps_{\mu\nu_2\nu_3\ldots\nu_d}\eps_{\rho_1\rho_2\ldots\rho_d}=\eps_{\rho_1\nu_2\nu_3\ldots\nu_d}\eps_{\mu\rho_2\ldots\rho_d}+\eps_{\rho_2\nu_2\nu_3\ldots\nu_d}\eps_{\rho_1\mu\rho_3\ldots\rho_d}+\ldots\nonumber\\
\ldots+\eps_{\rho_d\nu_2\nu_3\ldots\nu_d}\eps_{\rho_1\rho_2\ldots\rho_{d-1}\mu}\label{epid}
\eea
It amounts to swapping the index $\mu$ from the first epsilon with each index from the second epsilon.
\subsection*{Lagrangians and Equations of Motion}

Consider the Lagrangian for $n$ fields $\phi^i$ in $d$ space-time dimensions $x^\mu$ which does not involve a square root.
\be
{\cal L}= {\mathrm det}\left|\frac{\pd\phi^i}{\pd x_\mu}\frac{\pd\phi^j}{\pd x_\mu}\right|
\ee
The equations of motion for this Lagrangian are
\be
\frac{\pd^2 {\cal L}}{\pd \phi^i_\mu \pd \phi^j_\nu} \phi^j_{\mu \nu} = 0 \label{eomnsq}
\ee
These determinantal Lagrangians can be written as the sum of squares of Jacobians. The Jacobians will be denoted as
\be 
J_{\vkap}=J_{\kappa_1\kappa_2\ldots\kappa_{d-n}}=\eps_{\kappa_1\kappa_2\ldots\kappa_{d-n}\nu_1\nu_2\ldots\nu_n}\phi^1_{\nu_1}\phi^2_{\nu_2}\ldots\phi^n_{\nu_n}
\ee
For the square root case the Lagrangian is
\be
{\cal L}= \sqrt{{\mathrm det}\left|\frac{\pd\phi^i}{\pd x_\mu}\frac{\pd\phi^j}{\pd x_\mu}\right|}=\sqrt{\frac{1}{(d-n)!}J_{\vkap} J_{\vkap}}\label{lagsq}
\ee
The equations of motion for this can be written as
\be
J_{\mu\vkp} J_{\nu\vkp} \phi^i_{\mu \nu}=0\label{eomsq}
\ee

\subsection*{The Constraints}
The equations of motion for the non-square root case will be subject to the following constraints.
\be
\frac{\pd{\cal L}}{\pd \phi^i_\mu} \phi^i_{\mu \nu}=0. \label{const}
\ee
There is no summation over the index $i$.
The Lagrangian must also vanish.

The idea is to reduce the number of dimensions from $d$ to $d-1$. The constraints (\ref{const}) can be used to eliminate all second derivatives of the fields which involve a partial derivative with respect to $x_d$, the $d$th dimension. i.e From the constraints
\bea
\phi^i_{d \beta} = - \frac{\frac{\pd {\cal L}}{\pd \phi^i_\alpha}}{\frac{\pd {\cal L}}{\pd \phi^i_d}} \phi^i_{\alpha \beta}, \qquad
\phi^i_{dd} = \frac{\frac{\pd {\cal L}}{\pd \phi^i_\alpha}\frac{\pd {\cal L}}{\pd \phi^i_\beta}}{\left(\frac{\pd {\cal L}}{\pd \phi^i_d}\right)^2} \phi^i_{\alpha \beta}
\eea
Putting these into the equations of motion (\ref{eomnsq}) we have:
\bea
\sum_{j=1}^n\frac{1}{\left(\frac{\pd {\cal L}}{\pd\phi^j_d}\right)^2}\Biggl[\left(\frac{\pd {\cal L}}{\pd\phi^j_d}\right)^2 \frac{\pd^2 {\cal L}}{\pd\phi^i_\alpha \pd\phi^j_\beta}-\frac{\pd {\cal L}}{\pd \phi^j_d} \frac{\pd {\cal L}}{\pd \phi^j_\beta}\frac{\pd^2 {\cal L}}{\pd\phi^i_\alpha \pd\phi^j_d}
-\frac{\pd {\cal L}}{\pd \phi^j_d} \frac{\pd {\cal L}}{\pd \phi^j_\alpha}\frac{\pd^2 {\cal L}}{\pd\phi^i_\beta \pd\phi^j_d}\nonumber\\+\frac{\pd {\cal L}}{\pd\phi^j_\alpha}\frac{\pd {\cal L}}{\pd\phi^j_\beta} \frac{\pd^2 {\cal L}}{\pd\phi^i_d \pd\phi^j_d}\Biggr]\phi^j_{\alpha \beta}=0 \label{pr1}
\eea
The indices $\alpha,\beta=1,2,\ldots,(d-1)$ throughout the paper.

\subsection*{The Proof}
For the moment we shall consider the equation of motion with respect to field $\phi=\phi^1$ and are only looking at the component which involves the terms $\phi_{\alpha \beta}$. The other components will work in the same way. 
Using the definition of the Lagrangian which involves the Jacobians then we can write
\be
{\cal L} = \frac{1}{(d-n)!}\phi_\nu \phi_\rho B_{\nu\rho} \qquad {\mathrm where} \qquad  B_{\nu\rho}=\eps_{\nu\vkap\vnu}\eps_{\rho\vkap\vrho} \Phi_{\vnu} \Phi_{\vrho}
\ee
so the numerator of the coefficient of $\phi_{\alpha \beta}$ in (\ref{pr1}) becomes
\be
[B_{\mu d} (B_{\nu d} B_{\alpha \beta}-B_{\nu \beta} B_{\alpha d})+B_{\nu \alpha} (B_{\mu \beta} B_{dd}-B_{\mu d} B_{\beta d})]\phi_\mu \phi_\nu\label{pr2}
\ee
Now, 
\bea
B_{\nu d}B_{\alpha \beta}-B_{\nu \beta}B_{\alpha d}&=&[\eps_{\nu\vkap\vmu}\eps_{d\vkap\vnu}\eps_{\alpha\vtau\vrho}\eps_{\beta\vtau\vsig}-\eps_{\nu\vkap\vmu}\eps_{\beta\vkap\vnu}\eps_{\alpha\vtau\vrho}\eps_{d\vtau\vsig}]\ \Phi_{\vmu}\Phi_{\vnu}\Phi_{\vrho}\Phi_{\vsig}\nonumber\\
&=&\eps_{\nu\vkap\vmu}\eps_{\alpha\vtau\vrho}[\eps_{d\vkap\vnu}\eps_{\beta\vtau\vsig}-\eps_{\beta\vkap\vnu}\eps_{d\vtau\vsig}]\Phi_{\vmu}\Phi_{\vnu}\Phi_{\vrho}\Phi_{\vsig}\label{pr5}
\eea
Using the epsilon identity (\ref{epid}) to move the index $\beta$ around
\bea
\eps_{d\vkap\vnu}\eps_{\beta\vtau\vsig}=\eps_{\beta\vkap\vnu}\eps_{d\vtau\vsig}+\eps_{d\beta\kappa_2\ldots\kappa_r\vnu}\eps_{\kappa_1\vtau\vsig}+\ldots+\eps_{d\kappa_1\ldots\kappa_{r-1}\beta\vnu}\eps_{\kappa_r\vtau\vsig}+\nonumber\\\eps_{d\vkap\beta\nu_3\ldots\nu_n}\eps_{\nu_2\vtau\vsig}+\ldots+\eps_{d\vkap\nu_2\ldots\nu_{n-1}\beta}\eps_{\nu_n\vtau\vsig}\label{pr3}
\eea
The first term on the right hand side is just the other term in expression (\ref{pr5}). The last $(n-1)$ terms will all vanish due to symmetry conditions. This only leaves the middle terms. But by relabelling
\bea
\eps_{\nu \kappa_1\ldots\kappa_r\vmu}\eps_{d\kappa_1\ldots\beta\ldots\kappa_r\vnu}\eps_{\kappa_i\vtau\vsig}&=&\eps_{\nu \kappa_i\ldots\kappa_1\ldots\kappa_r\vmu}\eps_{d\kappa_i\ldots\beta\ldots\kappa_r\vnu}\eps_{\kappa_1\vtau\vsig}\nonumber\\
&=&\eps_{\nu\vkap\vmu}\eps_{d\beta\kappa_2\ldots\kappa_r\vnu}\eps_{\kappa_1\vtau\vsig}
\eea
There are $r=d-n$ of these terms. Therefore,
\be
B_{\nu d}B_{\alpha \beta}-B_{\nu \beta}B_{\alpha d}=r\eps_{\nu\vkap\vmu}\eps_{\alpha\vtau\vrho} \eps_{d\beta\vkp\vnu}\eps_{\kappa_1\vtau\vsig}\Phi_{\vmu}\Phi_{\vnu}\Phi_{\vrho}\Phi_{\vsig}
\ee  
Now as in (\ref{pr3}), using identity (\ref{epid}) to swap subscript $\kappa_1$
\bea
\eps_{\nu\vkap\vmu}\eps_{\alpha\vtau\vrho}&=&\eps_{\nu\alpha\kappa_2\ldots\kappa_r\vmu}\eps_{\kappa_1\vtau\vrho}+\eps_{\nu\tau_1\kappa_2\ldots\kappa_r\vmu}\eps_{\alpha\kappa_1\tau_2\ldots\tau_r\vrho}\ldots+\eps_{\nu\tau_r\kappa_2\ldots\kappa_r\vmu}\eps_{\alpha\tau_1\ldots\tau_{r-1}\kappa_1\vrho}\nonumber\\&&+\eps_{\nu\rho_2\kappa_2\ldots\kappa_r\vmu}\eps_{\alpha\vtau\kappa_1\rho_3\ldots\rho_n}\ldots+\eps_{\nu\rho_n\kappa_2\ldots\kappa_r\vmu}\eps_{\alpha\vtau\rho_2\ldots\rho_{n-1}\kappa_1}
\eea
And by relabelling indices and using the antisymmetric property of the epsilons
\be
\eps_{\nu\tau_i\kappa_2\ldots\kappa_n\vmu}\eps_{\alpha\tau_1\ldots\kappa_1\ldots\tau_r\vrho}\eps_{\kappa_1\vtau\vsig}=-\eps_{\nu\vkap\vmu}\eps_{\alpha\vtau\vrho}\eps_{\kappa_1\vtau\vsig}
\ee
so,
\be
(1+r) \eps_{\nu\vkap\vmu}\eps_{\alpha\vtau\vrho}\eps_{\kappa_1\vtau\vsig}=\eps_{\nu\alpha\vkp\vmu} \eps_{\kappa_1\vtau\vrho}\eps_{\kappa_1\vtau\vsig}
\ee
which gives
\be
B_{\nu d}B_{\alpha \beta}-B_{\nu \beta}B_{\alpha d}=\frac{r}{r+1} B_{\kappa\kappa}[\eps_{\nu\alpha\vkp\vmu} \eps_{d\beta\vkp\vnu} \Phi_{\vmu}\Phi_{\vnu}]
\ee
Substituting this into the expression (\ref{pr2}) we find
\bea
&&B_{\tau\tau}[B_{\mu d}\eps_{\nu\alpha\vkp\vmu}\eps_{d\beta\vkp\vnu}+B_{\nu\alpha}\eps_{\mu d\vkp\vmu}\eps_{\beta d\vkp\vnu}]\Phi_{\vmu}\Phi_{\vnu}\phi_\mu\phi_\nu\nonumber\\
&=&B_{\tau\tau}[\eps_{\mu\vtau\vrho}\eps_{d\vtau\vsig}\eps_{\nu\alpha\vkp\vmu}\eps_{d\beta\vkp\vnu}-\eps_{\mu\vtau\vrho}\eps_{\alpha\vtau\vsig}\eps_{\nu d\vkp\vmu}\eps_{d\beta\vkp\vnu}]\Phi_{\vmu}\Phi_{\vnu}\Phi_{\vrho}\Phi_{\vsig}\phi_\mu\phi_\nu
\eea
Now, using (\ref{epid}) to move the subscript $d$,
\bea
\eps_{d\vtau\vsig}\eps_{\nu\alpha\vkp\vmu}-\eps_{\alpha\vtau\vsig}\eps_{\nu d\vkp\vmu}&=&\eps_{\nu\vtau\vsig}\eps_{d\alpha\vkp\vmu}+\eps_{\kappa_2\vtau\vsig}\eps_{\nu\alpha d \kappa_3\ldots\kappa_n\vmu}+\ldots\nonumber\\&+&\eps_{\kappa_r\vtau\vsig}\eps_{\nu\alpha\kappa_2\ldots\kappa_{r-1}d\vmu}+\eps_{\mu_2\vtau\vsig}\eps_{\nu\alpha\vkpp d \mu_3\ldots\mu_n}\\&+&\ldots+\eps_{\mu_n\vtau\vsig}\eps_{\nu\alpha\vkpp \mu_2\ldots\mu_{n-1}d}\nonumber
\eea
As  before the last terms will vanish due to symmetry considerations. For the middle terms, by relabelling and using antisymmetry
\be
\eps_{\kappa_i\vtau\vsig}\eps_{\nu\alpha\kappa_2\ldots d\ldots\kappa_r\vmu}\eps_{d \beta\kappa_2\ldots\kappa_i\ldots\kappa_r\vnu}=\eps_{\kappa_2\vtau\vsig}\eps_{\nu\alpha d\vkpp\vmu}\eps_{d \beta\vkp\vnu}
\ee
There are $(r-1)=(d-n-1)$ of these terms. We now have
$$
\frac{r}{r+1}B_{\tau\tau}[\eps_{\mu\vtau\vrho}\eps_{\nu\vtau\vsig}\eps_{d\alpha\vkp\vmu}\eps_{d\beta\vkp\vnu}+(r-1)\eps_{\mu\vtau\vrho}\eps_{\kappa_2\vtau\vsig}\eps_{\nu\alpha d\vkpp\vmu}\eps_{d \beta\vkp\vnu}]\Phi_{\vmu}\Phi_{\vnu}\Phi_{\vrho}\Phi_{\vsig}\phi_\mu\phi_\nu
$$
Again, rewriting the epsilons, this time moving the subscript $\kappa_2$,
\bea
\eps_{\mu\vtau\vrho}\eps_{d \beta\vkp\vnu}&=&\eps_{\kappa_2\vtau\vrho}\eps_{d \beta\mu\vkpp\vnu}+\eps_{\mu\kappa_2\tau_2\ldots\tau_r\vrho}\eps_{d \beta\tau_1\vkpp\vnu}+\ldots+\eps_{\mu\tau_1\ldots\tau_{r-1}\kappa_2\vrho}\eps_{d \beta\tau_r\vkpp\vnu}\nonumber\\&&+\ldots+\eps_{\mu\vtau\kappa_2\rho_3\ldots\rho_n}\eps_{d\beta\rho_2\vkpp\vnu}+\ldots+\eps_{\mu\vtau\rho_2\ldots\rho_{n-1}\kappa_2}\eps_{d\beta\rho_n\vkpp\vnu}
\eea
And again, by relabelling
\be
\eps_{\mu\tau_1\ldots\kappa_2\ldots\tau_r\vrho}\eps_{d \beta\tau_i\vkpp\vnu}\eps_{\kappa_2\vtau\vsig}=-\eps_{\mu\vtau\vrho}\eps_{d\beta\vkp\vnu}\eps_{\kappa_2\vtau\vsig}
\ee
There are $r=d-n$ of these terms. 
So our expression is now
\bea
&\frac{r}{r+1}B_{\tau\tau}[\eps_{\mu\vtau\vrho}\eps_{\nu\vtau\vsig}\eps_{d\alpha\vkp\vmu}\eps_{d\beta\vkp\vnu}+\frac{r-1}{r+1}\eps_{\kappa_2\vtau\vrho}\eps_{\kappa_2\vtau\vsig}\eps_{\nu\alpha d\vkpp\vmu}\eps_{d \beta\vkpp\vnu}]\Phi_{\vmu}\Phi_{\vnu}\Phi_{\vrho}\Phi_{\vsig}\phi_\mu\phi_\nu\phi_{\alpha\beta}\nonumber\\
&=\frac{r}{r+1}B_{\tau\tau}[\eps_{d\alpha\vkp\vmu}\eps_{d\beta\vkp\vnu}\Phi_{\vmu}\Phi_{\vnu}r! {\cal L}-\frac{r-1}{r+1}B_{\kappa\kappa}J_{d\alpha\vkpp} J_{d\beta\vkpp}]\phi_{\alpha\beta}\nonumber\\
&=\frac{r r!}{r+1}\left(\frac{\pd J_{\vmu}}{\pd \phi_\tau}\frac{\pd J_{\vmu}}{\pd \phi_\tau}\right)\left[\left(\frac{\pd J_{d\vkp}}{\pd \phi_\alpha}\frac{\pd J_{d\vkp}}{\pd \phi_\beta}\right) {\cal L}-\frac{r-1}{(r+1)!}\left(\frac{\pd J_{\vnu}}{\pd \phi_\kappa}\frac{\pd J_{\vnu}}{\pd \phi_\kappa}\right) J_{d\alpha\vkpp}J_{d\beta\vkpp} \right]\phi_{\alpha\beta}\nonumber
\eea
A very similar calculation can be carried out to rewrite the coefficients of $\phi^j_{\alpha\beta}$ $(j\neq 1)$  from (\ref{pr1}). These become, for $\phi^j=\psi$, say.
\be
\frac{r r!}{r+1}\left(\frac{\pd J_{\vmu}}{\pd \phi_\tau}\frac{\pd J_{\vmu}}{\pd \psi_\tau}\right)\left[\left(\frac{\pd J_{d\vkp}}{\pd \psi_\alpha}\frac{\pd J_{d\vkp}}{\pd \psi_\beta}\right) {\cal L}-\frac{r-1}{(r+1)!}\left(\frac{\pd J_{\vnu}}{\pd \psi_\kappa}\frac{\pd J_{\vnu}}{\pd \psi_\kappa}\right) J_{d\alpha\vkpp}J_{d\beta\vkpp} \right] \psi_{\alpha\beta}
\ee
When the condition that the Lagrangian vanishes is put into the equations of motion, they can just be written as
\be
J_{d\alpha\vkpp} J_{d\beta\vkpp} \phi^i_{\alpha\beta}=0\label{ans}
\ee
as required.
Comparing (\ref{ans}) with (\ref{eomsq}), these are the equations of motion for the Lagrangian involving a square root (\ref{lagsq}) in $(d-1)$ dimensions.

\end{document}